\documentclass[aps,prd, twocolumn, nofootinbib, 10pt, preprintnumbers]{revtex4}
\usepackage{graphicx}
\usepackage{epstopdf}
\usepackage{graphicx, epsfig}
\usepackage{color}
\usepackage{mathrsfs}
\usepackage{amsmath}
\usepackage{amssymb}
\usepackage{hyperref}

\newcommand{\beq}{\begin{equation}}
\newcommand{\eeq}{\end{equation}}
\newcommand{\bea}{\begin{eqnarray}}
\newcommand{\eea}{\end{eqnarray}}

\newcommand{\ben}{\begin{enumerate}}
\newcommand{\een}{\end{enumerate}}

\newcommand{\be}{\begin{equation}}
\def\bel#1{\begin{equation} \label{#1}}
\newcommand{\ee}{\end{equation}}
\newcommand{\bi}{\begin{itemize}}
\newcommand{\ei}{\end{itemize}}
\newcommand{\ba}{\begin{align}}
\newcommand{\ea}{\end{align}}
\newcommand{\comments}[1]{}

\begin{document}
\title{Viability of Arctan Model of $f(R)$ Gravity for Late-time Cosmology}
\author{Koushik Dutta\footnote{email:koushik.dutta@saha.ac.in},
       }
\affiliation{Theory Division,
Saha Institute of Nuclear Physics,\\ 1/AF Salt Lake,
Kolkata - 700064, India.}

\author{Sukanta Panda\footnote{email:sukanta@iiserb.ac.in},
        Avani Patel\footnote{email: avani@iiserb.ac.in}}
        \affiliation{ Indian Institute of Science Education and Research,\\ Bhauri, Bhopal 462066, Madhya Pradesh, India}
\begin{abstract}
$f(R)$ modifications of Einstein's gravity is an interesting possibility to explain the late time acceleration of the Universe. In this work we explore the cosmological viability of one such $f(R)$ modification proposed in \cite{Kruglov:2013qaa}. We show that the model violates fifth-force constraints. The model is also plagued with the issue of curvature singularity in a spherically collapsing object, where the effective scalar field reaches to the point of diverging scalar curvature.
\end{abstract}
\maketitle

\section{Introduction}
Last two decades have seen substantial improvement in understanding the large-scale structures of the universe from high precision measurements of cosmic microwave background radiation and distance measurements of Type Ia Supernovae \cite{Nobel2011}. The later observations have led us to conclude that the universe we live in is expanding acceleratingly in recent times. In order to accommodate such evolution of the universe, the standard model of cosmology is thought with a cosmological constant term, $\Lambda$. This is the simplest extension to the Einstein-Hilbert action. Moreover, it is consistent with all available cosmological data till date and is commonly known as $\Lambda$ cold dark matter ($\Lambda$-CDM) model. However, it is very difficult to explain the origin of the required value of the cosmological constant from any fundamental physics \cite{Padilla:2015aaa}. 

As an alternative to the $\Lambda$-CDM model, modifications of gravity action by higher order curvature invariant terms are considered. The most promising models in this category are f(R) theories of gravity \cite{f(R)TheoriesReview}. 
In an $f(R)$-theory Lagrangian, the Ricci scalar $R$ is replaced by an analytical function $f(R)$. Initially, the diverging $f(R)$ models at $R=0$, e.g. inverse-power law models, $ f(R) \propto R^n$ with $n<0$, were proposed for late-time acceleration \cite{Carroll:2003wy}. But, the models were shown to be unviable because of matter instability \cite{Dolgov:2003px} and failure to provide matter era before the accelerating phase \cite{Amendola:2006we}. Additionally, the models also violate the fifth-force constraints by carrying long range force provided by extra scalar d.o.f in the theory \cite{Chiba:2006jp}.  All these above-mentioned issues severely constrain the allowed form of $f(R)$, and it has steered the streamlining of $f(R)$ models towards a new class of models which are analytical at $R=0$. 

Various observational data are putting the possibility of a cosmological constant as the best fit dark energy model. Following this line, many $f(R)$ models are proposed those behave as $\Lambda$-CDM model when the spacetime is sufficiently curved $i.e.$ $f\rightarrow R-2\Lambda$ for $R\gg\Lambda$ and $f(0)=0$ \cite{Hu:2007nk,Starobinsky:2007hu,Appleby:2007vb,Appleby:2009uf,Tsujikawa:2007xu,Zhang:2005vt,Cognola:2007zu,Miranda:2009rs,Linder:2009jz,Bamba:2010ws,Bamba:2010zz,Kruglov:2013qaa}. The dynamical system analysis of $f(R)$ theories is carried out in categorising  models according to their fixed points \cite{Amendola:2006we}. Fifth force constraints on these $f(R)$ models are evaded using the chameleon mechanism \cite{Khoury:2003aq, Capozziello:2007eu}.  

One of the serious problems in these models is the occurrence of singularities of various types. It was observed that the minimum of the scalar field potential can be near to the singularity point ($R\rightarrow\infty$), and hence it is likely that the scalar field hits the singularity if the model parameters are not fine tuned appropriately \cite{Appleby:2008tv, Frolov:2008uf}. Since the potential well becomes shallower in the presence of matter density, the possibility of the occurrence of singularity increases  in a matter distribution \cite{Frolov:2008uf}.  The occurrence of curvature singularity can also be seen in a collapsing astrophysical object. In this case, the singularity is analysed for suitable $f(R)$ models applied to dense objects undergoing contraction in the presence of linearly time-dependent collapsing mass density \cite{Arbuzova:2010iu, Lee:2012dk, Reverberi:2012ew}. It is seen that the singularity is reached in a time that is much shorter than the cosmological time scale. In \cite{Dutta:2015nga}, both static and dynamical analysis in the contracting astrophysical object is carried out for a general $f(R)$ model proposed in \cite{Miranda:2009rs}. It was found that the models that satisfy the fifth-force constraints are typically plagued with the curvature singularity issue. It is also noted that the issue of curvature singularity can be eliminated by adding an extra curvature term to the Lagrangian \cite{Appleby:2008tv, Appleby:2009uf}. The finite-time singularity in modified gravity is also described in \cite{Nojiri:2008fk, Bamba:2008ut}. It is shown that the past singularities may be prevented for a certain range of parameters. These singularities may also occur in future and can be avoided for fine-tuned initial conditions \cite{Dev:2008rx, Thongkool:2009js}. 

In this work, our primary aim is to examine the viability of the model proposed in \cite{Kruglov:2013qaa}. We will consider fifth-force constraint analysis, and also investigate the existence of curvature singularity along the line of \cite{Dutta:2015nga}. Additionally, we also carry out the dynamical system analysis for the model in pointing out the differences with \cite{Kruglov:2013qaa}. In the above work, other than the late-time cosmology, inflationary dynamics were also investigated. In our work, we will concentrate only on the viability of this model for late-time cosmology.  This work is organised as follows: Sec.~\ref{sec:arctan} gives the general idea about the ArcTan model and its de Sitter points. Sec.~\ref{subsec:fixedpoints} discusses the fixed points of the model in understanding its proper cosmological evolution as a late time dark energy model. The fifth-force constraints are analysed in Sec.~\ref{sec:fifthforce}. The investigation of curvature singularity is carried out in Sec.~\ref{sec:cursingularity}, with a conclusion in the final Sec.~\ref{sec:conclusion}.

\section{$f(R) = tan^{-1} R$ Model}\label{sec:arctan}
As a modified theory of gravity, $f(R)$ theory is described by the following action 
\begin{equation}
S=\int d^4x \sqrt{-g}\left[\frac{1}{2\kappa^2} f(R)+\mathfrak{L}_m\right],
\label{action}
\end{equation}
where $f(R)$ is an arbitrary function of Ricci scalar $R$, and it is written in the following form where we separate out the usual Einstein Hilbert contribution: $f(R)=R + F(R)$. $\mathfrak{L}_m$ is the matter part of the Lagrangian. As mentioned earlier, the function $f(R)$ given in Eq.~\eqref{action} satisfies the condition $f(0) = 0$, and $f(R) \rightarrow R - 2\Lambda$ at high curvature so that the early-time cosmology is identical to $\Lambda$-CDM cosmology and physics is modified at the infrared scales (latetime) only. 
By varying the action w.r.t. $g_{\mu\nu}$ we obtain the field equation whose trace is given by
\begin{equation}
3\Box F_{,R}(R)-2F-R+RF_{,R}(R)=\kappa^2 T ~,
\label{trace}
\end{equation}
where a comma in the subscript denotes derivative w.r.t to Ricci scalar. We note that non-vanishing $\Box F_{,R}$ term gives an extra dynamical scalar degree of freedom $\phi=F_{,R}$ other than the usual graviton. 

A new model of $f(R)$ theory has been proposed in \cite{Kruglov:2013qaa} where the function $F(R)$ is given by
\begin{equation}
F(R)=-\frac{b}{\beta}tan^{-1}(\beta R)~,
\label{model}
\end{equation} 
with $\beta$ being positive and having inverse mass dimension two.  Note that the model considered here in Eq.\eqref{model} reduces to $\Lambda$-CDM model in high curvature regime i.e. $R\gg1/\beta$.
The first and second derivatives of $f(R)$ w.r.t $R$ for this model are given by
\begin{equation}
f_{,R}(R)=1-\frac{b}{1+( \beta R)^2},~~~~f_{,RR}(R)=\frac{2b\beta^2 R}{\left(1+(\beta R)^2\right)^2}.
\label{3}
\end{equation}
The condition for scalar field $\phi$ to be non-tachyonic ($f_{,RR} > 0$) requires $b > 0$, and the condition for graviton to be of non-ghost nature ($f_{,R} > 0$) requires  
\begin{equation}
1+(\beta R)^2 > b, ~~{\text for}~~ R_d < R < \infty, 
\label{4}
\end{equation} 
with $b >0$. We thus take $0 < b < 1$ for all following considerations. Here, $R_d$ is the constant curvature solution of Eq.~\eqref{trace} for vacuum, and can be obtained by solving the following condition \cite{Kruglov:2013qaa} 
\begin{equation}
\frac{-bR_d}{1+(\beta R_d)^2}-R_d+\frac{2b}{\beta}tan^{-1}(\beta R_d)=0.
\label{condition}
\end{equation}
Note that fixing the value of $\beta R_d$ in Eq.~\eqref{condition}, one can fix the value of $b$. Other than the trivial solution of $R_0 = 0$ corresponding to the Minkowski space-time, there are two solutions of Eq.~\eqref{condition}, of which one is an unstable de Sitter point $x_1=\beta R_d^{(1)}$ and another is a stable de Sitter point $x_2=\beta R_d^{(2)}$. If a de Sitter point $R_d$ satisfies the condition $F_{,R}(R_d)/F_{,RR}(R_d)>R_d$, then it is stable and can describe primordial and the present epoch dominated by vacuum energy. For $b < 0.93$ the Eq.~\eqref{condition} has only trivial solution $R_0 = 0$. All the de Sitter points for the allowed values of $b$ are summarised in Table.~\ref{tabl:desitter}.
\begin{table}
\centering
\begin{tabular}{|l|l|l|r|}
  \hline
  $ b$           & $x_1$ & $x_2$ \\
  \hline
   0.93     &                     0.7304    & 0.9199    \\
   0.94      &                    0.5776    & 1.0821    \\
   0.95      &                    0.4852    & 1.1815     \\
   0.96       &                   0.4081    & 1.2624      \\
   0.97       &                   0.3361    & 1.3334      \\
   0.98       &                   0.2630    & 1.3980       \\
   0.99       &                   0.1791    & 1.4582       \\
  \hline
\end{tabular}
\caption{de Sitter points for different values of $b.$}
\label{tabl:desitter}
\end{table}

\subsection{Viability as a Dark Energy Model}\label{subsec:fixedpoints}
Any $f(R)$ modification of late-time cosmology should give rise to accelerating expansion at the present epoch preceded by a matter dominated era. A general dynamical analysis of $f(R)$ theory is carried out in \cite{Amendola:2006we}, and cosmological viability conditions are derived. The field equations are rewritten in terms of a set of first order autonomous differential equations for dimensionless variables $y_1=-\dot{f_{,R}}/Hf_{,R}$, $y_2=-f/6f_{,R}H^2$, and $y_3=R/6H^2$, where following quantities were defined
\begin{equation}
m\equiv\frac{Rf_{,RR}}{f_{,R}}=\frac{2bx^2}{(1+x^2)(1-b+x^2)},
\end{equation}
\begin{equation}
r\equiv-\frac{Rf_{,R}}{f}=-\frac{x(1-b+x^2)}{(1+x^2)(x-b\arctan(x))}=\frac{y_3}{y_2}.
\end{equation}
Here, $x\equiv \beta R.$ There are six fixed points characterised by matter density $\Omega_m(m)$ and $w_{eff} (m) = -1 - 2 \dot H/(3 H^2)$. The point $P_5$ and $P_6$ fall on the line $m(r) = - r -1$, and the $\Lambda$-CDM cosmological evolution is denoted by $m = 0$ line. The point $P_5 (r\simeq-1,m\simeq +0)$ corresponds to matter dominated era with $w_{eff} \simeq 0$.  The points $P_1(r = -2, 0<m\leq 1)$ and $P_6(m = - r - 1,(\sqrt 3-1)/2<m<1)$ both correspond to accelerating expansion of the Universe, where the former one is a de Sitter point. Therefore, all viable dark energy models fall into two classes \cite{Amendola:2006we}: \\
Class II: Models that connect $P_5$ to $P_1$, Class IV: Models that connect $P_5$ to $P_6$.
For a particular model, $m$ can be plotted as a function of $r$, and its cosmological viability can be tested.

Dynamical analysis and stability of critical points for the model of Eq.~\eqref{model} have been studied in \cite{Kruglov:2013qaa}. Here, we reanalyse the stability of all the critical points.
\begin{figure}
 \centering
 \includegraphics[width=0.44\textwidth]{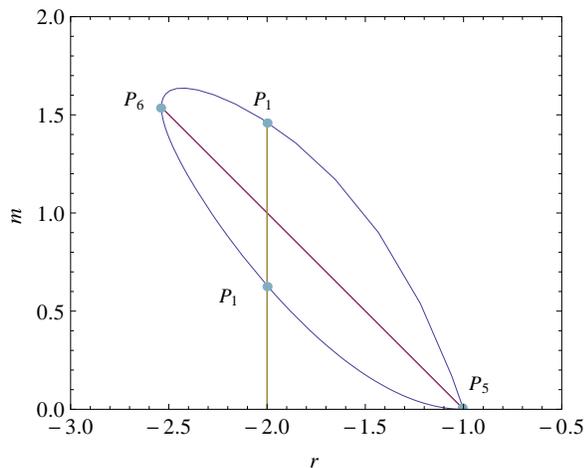}
 \caption{Trajectory in the $m-r$ plane for $b=0.99$. Fixed points are marked with blue dots.}
 \label{fig:fixedpoints}
\end{figure}
The $m$ vs $r$ plot is shown in Fig.\ref{fig:fixedpoints} for $b=0.99$, and all the critical points are marked. One important characteristic of this model compared to many previously analysed model is that the  plot is multivalued. The upper and lower branch is separated by the line $m = -r -1$. As we move clockwise along the curve the scalar curvature $R$ decreases with point $P_5 (r\sim-1,m\simeq +0)$ corresponding to both small and high curvature limits. 

The intersection of $m(r)$ curve with $r=-2$ line gives the de Sitter points $P_1$. We have two $P_1$ points at $x=0.1791$ and $x=1.4582$ for $b=0.99$. The stability condition for the stable de Sitter point is at $r=-2,0<m \leq 1$.  The de Sitter point $x=0.1791$ belongs to the upper branch of the curve. Since, at this de Sitter point, $m(r=-2)>1$, it is an unstable point. The point $x=1.4582$ corresponding to lower branch of the curve is a stable de Sitter point. 

The points $P_5$ and $P_6$ can be located at the intersections of the $m(r)$ (blue) curve with $m=-r-1$ (red) line. The point $P_6$ is located at $m\sim 1.54$ and $r  \simeq -2.54$ with $x\simeq 0.48$. As $m(r\simeq-2.54) >1$, the point $P_6$ is an unstable point. Point $P_5$ represents the saddle matter era. The condition for $P_5$ to exist is $m (r=-1)=0$. It is shown in \cite{Kruglov:2013qaa} that the point $P_5$ is situated at $x=\beta R\sim 0$. But, saddle matter point has to be at higher curvature than the de Sitter point $P_1$ to explain the cosmic history correctly. Moreover, $dm/dr$ at $P_5$ has to be greater than $-1$ for the existence of acceptable saddle matter era. From the definitions of $m$ and $r$ and Eq.\eqref{model}, one can obtain the expression of $m'(r)$ \cite{Kruglov:2013qaa}. 
Taking $x\rightarrow 0$ limit one can find that $m' (r=-1)\rightarrow-3<-1$. Thus the point $P_5$ with $x=0$ is not an acceptable point being unstable to its perturbations. On the other hand, in the limit of $x\gg1$ i.e. $R\gg1/\beta$, $r$ goes to $-1$ with $m$ approaching zero. One can find that $m'(r=-1)\rightarrow -0.0025 > -1$. Thus, in the large curvature limit the point $P_5$ is stable under perturbations. In conclusion, the model of Eq.~\eqref{model} gives saddle matter era at very large value of curvature and the Universe moves from $P_5$ to $P_1$ (in the bottom branch) in its cosmological evolution. According to the original reference of \cite{Amendola:2006we}, the model belongs to the Class II category of $f(R)$ models.

\section{Local Gravity Constraint}\label{sec:fifthforce}
We have seen in the previous section that when cosmological evolution happens from $P_5$ to $P_1$ via the lower branch of Fig.~\ref{fig:fixedpoints}, the present dark energy dominated epoch is preceded by the ordinary matter dominated era. But the form of $f(R)$ should not spoil the experimentally verified results of General Relativity at local scales. The fifth force originated by an extra scalar degree of freedom in an $f(R)$ theory must be attenuated on local gravitational systems like earth and solar system so that the theory can evade the local gravity tests. In the Einstein frame, the scalar field $\psi$ (corresponding to $\phi$ in Jordon frame) is a chameleon-like field which couples to the matter in such a way that the effective mass of the scalar field depends on the local matter density \cite{Khoury:2003aq}.

In the Einstein frame, the action can be rewritten as 
\begin{equation}
S = \int{d^4x \sqrt{-\tilde g}\left[\frac{\tilde R}{2\kappa^2} - \frac{(\tilde\nabla\psi)^2}{2} - V_E(\psi) + \mathfrak{L}_m(\tilde g_{\mu\nu}e^{-\frac{2}{\sqrt{6}}\kappa\psi})\right]},
\label{11}
\end{equation}
where all quantities having tilde are defined in the Einstein frame. 
The scalar field $\psi$ for the model of Eq.~\eqref{model}, in the high curvature regime (where $R \gg \frac{1}{\beta}$ and $F_{,R}\ll1$) is given by
\begin{equation}
\psi=\sqrt{\frac{3}{2\kappa^2}}\ln f_{,R}=\sqrt{\frac{3}{2\kappa^2}}\ln(1+F_{,R})\approx\sqrt{\frac{3}{2\kappa^2}} F_{,R}.
\label{scalarfieldein}
\end{equation}
The potential $V_E(\psi)$ is given by
\begin{eqnarray}
V_E(\psi) =\left.\frac{Rf_{,R}(R)-f(R)}{2\kappa^2f_{,R}^2(R)}\right|_{R=R(\psi)}=(1+(\beta R)^2) \times \nonumber\\
\left.\frac{\left[\left(1+(\beta R)^2\right)\frac{b}{\beta}tan^{-1}(\beta R)-b R\right]}{2\kappa^2\left[1-b+(\beta R)^2\right]^2}\right|_{R=R(\psi)},
\label{13}
\end{eqnarray}
where $R = R(\psi)$ needs to be substituted by inverting Eq.~\eqref{scalarfieldein}. 

\begin{figure}
 \centering
 \includegraphics[width=0.44\textwidth]{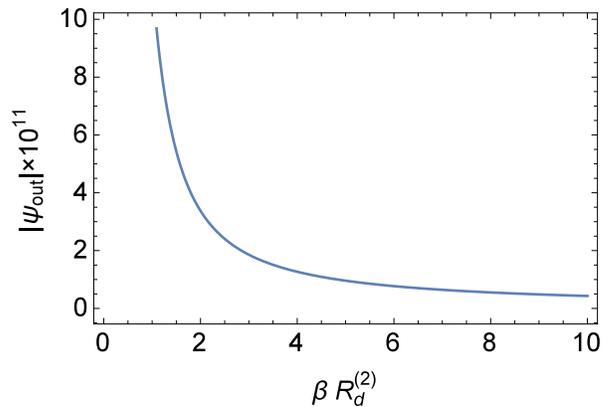}
 \caption{$|\psi_{out}|$ vs. $\beta R^{(2)}_d$.}
 \label{fig:fifthforcearctan}
\end{figure}

Let us consider a spherically symmetric body with radius $\tilde r_c$. 
The effective potential $V_{\text {eff}}$ is defined by 
\begin{equation}
V_{\text{eff}}(\psi)=V(\psi)+e^{-\frac{2}{\sqrt 6}\kappa\psi}\rho^*~,
\label{effpotein}
\end{equation}
where $\rho^*$ is a conserved quantity in the Einstein frame. We assume that the spherically symmetric body has a constant density $\rho^*=\rho_{in}$ inside the body ($\tilde{r}<\tilde{r}_c$) and $\rho^*=\rho_{out}(\ll\rho_{in})$ outside ($\tilde{r}>\tilde{r}_c$). $\psi_{in}$ and $\psi_{out}$ are the values of the field at the minima of the effective potential $V_{\text{eff}}$ inside and outside the object respectively. 
The thin shell parameter is given by \cite{Khoury:2003aq}
\begin{equation}
\frac{\delta\tilde{r}_c}{\tilde{r}_c}=-\frac{\psi_{out}-\psi_{in}}{\sqrt 6\Phi_c}~,
\end{equation}
where $\Phi_c$ is the gravitational potential of the test body (Sun/Earth). This shows that the only thin shell having width $\delta\tilde r_c$ around the surface of the object contributes to the field outside the object thus resulting into the suppression of the fifth force on the surface of the test body.
Since $|\psi_{out}|\gg|\psi_{in}|$, the above equation reduces for our case to
\begin{equation}
\left|\psi_{out}\right|\simeq\sqrt{6}\Phi_c\frac{\delta\tilde{r}_c}{\tilde{r}_c}.
\end{equation}
To evade the local gravity tests, the right hand side of the above equation should be \cite{Will:2014xja,Capozziello:2007eu}\\
\begin{eqnarray}
\label{eq:phiconstraint}
&\lesssim& \left \{ \begin{array}{rl}
 5.97\times 10^{-11} &\textrm{(Solar system test)}, \\
 3.43\times 10^{-15} &\textrm{(Equivalence Principle test)}.
\end{array} \right .
\label{ffcond}
\end{eqnarray}
Using $R_{\text {out}} = \kappa \rho_{\text{out}}$, from Eq.\eqref{scalarfieldein} and Eq.\eqref{effpotein} we obtain
\begin{equation}
\left|\psi_{out}\right|\approx\frac{\sqrt{6}}{2\kappa}\frac{b}{(\beta\kappa\rho_{out})^2}=\frac{\sqrt{6}}{2\kappa}\frac{b}{((x_2/R_d^{(2)})\kappa\rho_{out})^2}.
\label{phib1}
\end{equation}
In the previous section, we have found that there exists stable de Sitter points only for $0.93\leqslant b<1$. For an example let us consider $b=0.97$ which has a stable de Sitter point at $x_2=\beta R_d^{(2)}=1.3334$. 
From the fact that the energy density of the baryonic/dark matter in our galaxy $\rho_{out}$ is $\sim 10^{-24} g/cm^3$ and the curvature at the de Sitter minimum is roughly of the order of $\rho_c \simeq10^{-29} g/cm^3, $ $\left|\psi_{out}\right|$ comes out to be
\begin{equation}
\left|\psi_{out}\right|\approx 6.682 \times 10^{-11}~.
\end{equation}
Since $|\psi_{out}|$ is large compared to the required values given in Eq.\eqref{ffcond}, we can say that this model does not evade the local gravity constraints for $b=0.97$. We can get the same result for all the values of $b$ in the acceptable range given in Table~\ref{tabl:desitter}. It can be seen from Eq.~\eqref{condition} that $b$ is not independent but varies as the value of $\beta R_d$ varies. In Fig.~\ref{fig:fifthforcearctan} we show $|\psi_{out}|$ with respect to $x_2(=\beta  R_d^{(2)})$. It is clearly seen that $|\psi_{out}|$ is greater than $0.5\times10^{-11}$. From this result, one can draw the conclusion that the model given by Eq.~\eqref{model} hardly satisfies the Solar System constraint and does not satisfy Equivalence Principle constraint.

\section{Curvature Singularity in Arctan Model}\label{sec:cursingularity}
In this section, we will analyse the behaviour of the effective scalar degree of freedom in a system of collapsing mass density. We will show that in finite time, the field evolves to a point where the Ricci scalar diverges. 

From Eq.~\eqref{trace}, we have seen that an $f(R)$ theory has an extra scalar degree of freedom $\phi=F_{,R}$, compared to General Relativity. The associated dynamics of the field is controlled by Eq.\eqref{trace}, and can be rewritten as
\begin{equation}
\Box\phi=\frac{dV_J}{d\phi}+\frac{\kappa^2}{3}T,
\label{oscillations}
\end{equation}
where
\begin{equation}
\frac{dV_J}{d\phi}=\frac{1}{3}(R+2F-RF_{,R}).
\label{dvjbydphi}
\end{equation}
The Eq.~\eqref{oscillations} corresponds to an oscillator where the energy-momentum part behaves as a force term. We prefer to work in the Jordan frame since it is more convenient to examine the issue of curvature singularity in the Jordan frame than in the Einstein frame. 

The oscillations of the scalar field $\phi$ are governed by the potential $V_J$, and the form of $V_J$ depends on the function $f(R)$ in a given model. Inverting the relation $\phi=F_{,R}$ to write the Ricci scalar $R$ in terms of $\phi$, and integrating Eq.~\eqref{dvjbydphi} w.r.t. $\phi$, we obtain the form of the potential $V_J$.  In vacuum, the field $\phi$ oscillates around the minimum $\phi_{min}$ of the potential which is also a de Sitter point \cite{Dutta:2015nga}. Cosmological evolution happens around this point. There is also a point $\phi_{sing}$ where Ricci scalar diverges $R \rightarrow \infty$, and it is finite field distance away from the minimum $\phi_{min}$. While the scalar field $\phi$ oscillates around $\phi_{min}$, it is energetically possible for the field to hit the singularity if the potential difference between $\phi_{min}$ and $\phi_{sing}$ is finite.

We first analyse whether the curvature singularity point exists in the model of Eq.~\eqref{model}, and secondly we investigate the evolution of the Ricci scalar in a collapsing object whose energy density is linearly growing. In presence of matter-energy density, the oscillations of the field $\phi$ are governed by the effective potential $V_J^{eff}$. In this case, the Eqns.\eqref{oscillations} and \eqref{dvjbydphi} can be rewitten as 
\begin{equation}
\Box\phi=\frac{\partial V_J^{eff}}{\partial\phi}~,
\label{vjeffective}
\end{equation}
\begin{equation}
\frac{\partial V_J^{eff}}{\partial\phi}=\frac{1}{3}(R+2F-RF_{,R}+\kappa^2T).
\end{equation}
In fact, the minimum of the potential $V_J^{eff}$ shifts from the de Sitter point, and moves closer to the curvature singularity point. Thus the effects of matter are necessary to be investigated even when a model is well behaved in vacuum \cite{Frolov:2008uf, Dutta:2015nga}. 

Let us first examine the profile of the potential $V_J$ for vacuum i.e, $\kappa^2T=0$. For our model, the scalar field $\phi$ is given by
\begin{equation}
\phi  = -\frac{b}{1+x^2}.
\label{scalarfield}
\end{equation}
Integrating Eq.~\eqref{dvjbydphi} w.r.t. $\phi$, we obtain the potential in terms of $\phi$ as
\begin{eqnarray}
\beta V_J=-\frac{\phi^2\sqrt{-1-b/\phi}}{6}+\frac{(4+3b)\phi\sqrt{-1-b/\phi}}{12}  \nonumber\\
-\frac{2b}{3}\left(\frac{1}{8}(-4+3b)+\phi\right)tan^{-1}(\sqrt{-1-b/\phi})~.
\label{jordanpot1}
\end{eqnarray}
From Eq.~\eqref{scalarfield}, it is clear that as $\phi\rightarrow 0$, $x\rightarrow\infty$, leading to a curvature singularity. 

We now probe the height of the potential barrier between the de Sitter minimum $\phi_{min}$ and the singular point $\phi_{sing}$. In the region between $\phi_{min}$ and $\phi_{sing}$, $-b/\phi$ varies from $1+(\beta R_d)^2$ to $\infty$, and therefore we can take the limit $-b/\phi\gg1$ for this region in Eq.~\eqref{jordanpot1}. It can be easily seen that the third term of the potential in Eq.\eqref{jordanpot1} dominates over other terms in this limit, and the Eq.~\eqref{jordanpot1} can be rewritten as
\begin{equation}
\beta V_J=-\frac{b}{12}(-4+3b)tan^{-1}\left(\sqrt{- b/\phi}\right)~.
\end{equation}
Since $tan^{-1}(\sqrt{-b/\phi})$ goes to a finite constant value for large $\sqrt{-b/\phi}$, we have $\beta V_J\propto b$. We thus conclude that the height of the potential barrier is finite and proportional to the value of $b$. This makes the model \eqref{model} vulnerable to curvature singularity.
\begin{figure}
 \centering
 \includegraphics[width=0.44\textwidth]{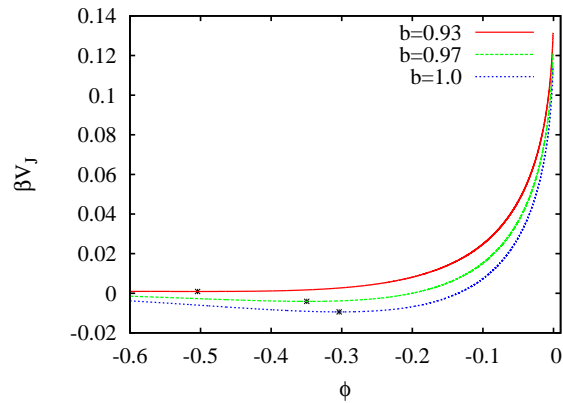}
 \caption{$\beta V_J$ vs. $\phi$ for different values of parameter $b$. The de Sitter points are marked by $*$ in the plot.}
 \label{fig:desitterandsingularity}
\end{figure}
In Fig.~[\ref{fig:desitterandsingularity}], potential $\beta V_J$ vs. $\phi$ has been plotted for different values of the parameter $b$. $\phi=0$ corresponds to the curvature singularity point where curvature $R$ diverges to infinity. The de Sitter points are marked by ``$*$" in the plot. It can be seen that the de Sitter points for larger values of $b$ are at a greater depth from the singularity causing the lesser probability of scalar field $\phi$ reaching singularity. 

To make our analysis more quantitative we need to solve the equation of motion and confirm that curvature scalar $R$ indeed diverges. For this, we study the issue in an astro-physical object like galactic cloud of dust which collapses under its own gravity. Acknowledging the fact that its density increases with time, the energy-momentum tensor for such a system can be empirically taken as
\begin{equation}
T=-\kappa^2T_0(1+t/t_{ch}).
\end{equation}
Here, $t_{ch}$ is the characteristic time of the collapsing object. The value of typical characteristic time for a collapsing galactic cloud is $t_{ch} \sim1.34\times 10^{15} sec$ \cite{Dutta:2015nga}. Though the above form of $T$ is not exact, it can give qualitatively correct scenario, provided that the contraction is slow enough. Since the density of dust cloud, $\rho_m=10^{-24} g/cm^3$ is much greater than the average density of the universe $i.e.$ $\rho_{crit}=10^{-29}g/cm^3$, we can take the limit $R \gg1/\beta$ in Eq.~\eqref{trace}. Nevertheless the astrophysical density $\rho_m$ is low enough to consider the background metric as a Minkowski metric \cite{Reverberi:2012ew}. Therefore, we can replace covariant derivatives with partial derivatives in Eq.~\eqref{trace}. Moreover, spatial derivatives are also ignored because of the presumed homogeneity and isotropy. 

Considering new variables $y=\kappa^2T_0/R$ and $\tau=t/t_{ch}$, the equation of motion can be obtained in terms of $y$ and $\tau$ as
\begin{eqnarray}
y''+\frac{{y'}^2}{y}+\tau_{ch}^2y^{-1}\left[\frac{1}{3}\left(1+\frac{\tau}{\tau_{ch}}\right)-\frac{y^{-1}}{3}+ \right.  \nonumber \\
\frac{2}{3\alpha}\left.\tan^{-1}(\alpha y^{-1})-\frac{by}{3\alpha^2}\right]=0~,
\label{dyneqiny}
\end{eqnarray}
where $\alpha=\beta\kappa^2T_0$, $\tau_{ch}=\sqrt{\alpha^3/2b\beta}\,t_{ch}$, and prime denotes derivative w.r.t. $\tau$. We solve the above equation and inquire what happens to $y(\tau)$ within characteristic time as the object collapses.
\begin{figure}
 \centering
 \includegraphics[width=0.44\textwidth]{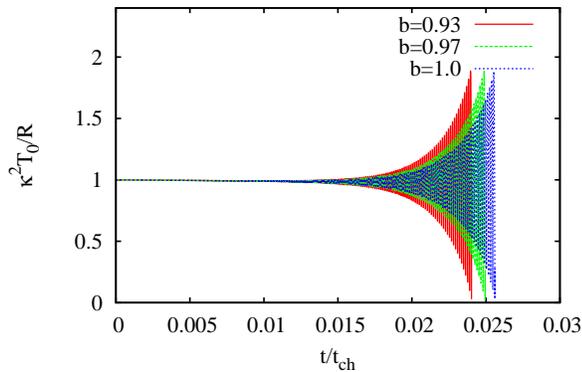}
 \caption{Oscillations of $y=\kappa^2T_0/R$ vs. $t/t_{ch}$.}
 \label{fig:dynsolution}
\end{figure} 
Here, $\alpha\sim\rho_m/\rho_{crit}\sim 10^5$ and $\beta$ is average curvature of the universe at present time and its numerical value is given by $\beta=1/t_U^2$, where $t_U$ is the age of the universe. 

To solve Eq.~\eqref{dyneqiny}, we take $y'(0)=0$ and $y(0)=1$ as initial conditions. The solutions of Eq.~\eqref{dyneqiny} are plotted in Fig.~[\ref{fig:dynsolution}] for $b=0.93,0.97$ and $1.0$. One can notice that $R\rightarrow\infty$ corresponds to $y\rightarrow 0$. The Fig.~[\ref{fig:dynsolution}] clearly shows that the oscillations of $y$ gradually increases and after a finite time it reaches to zero resulting into curvature singularity. It can be seen that $y$ becomes zero sooner for smaller $b$, and later for higher values of $b$. This is consistent with what we obtained by our previous static analysis by looking at the potential. From our numerical solutions of Eq.~\eqref{dyneqiny}, we obtain $t_{sing}=0.032\times 10^{15},\,0.033\times 10^{15}$ and $0.035\times 10^{15}\, sec$ for $b=0.93,\,0.97$ and $1.0$ respectively. Thus, we find that the collapsing object encounters the curvature singularity within time much smaller than the cosmological time scale.

\section{Conclusions}\label{sec:conclusion}
In this paper, we investigate cosmological viability of a model proposed in \cite{Kruglov:2013qaa} where $f(R) = tan^{-1} R$. We investigate the fifth force constraint and show that it immediately violates the observational tests. Following the line of \cite{Dutta:2015nga}, the issue of curvature singularity is investigated in the Jordan frame. 

To check the cosmological viability of the model, we do fixed point analysis and point out some differences with the results found in \cite{Kruglov:2013qaa}. The stable fixed point $P_6$ which is responsible to give rise to the accelerated expansion is not present in this model. We find the stable fixed point $P_5$ at very large value of the curvature. The point $P_5$ corresponds to a saddle matter era which is at curvature higher than the de Sitter era given by $P_1$. We also examine the viability of the model at local gravity scales by putting fifth force constraint through chameleon mechanism given in \cite{Khoury:2003aq}.

We find out that in the field space there exists a singularity point where the curvature scalar diverges to infinity. The potential barrier between the de Sitter minimum and the singularity point is finite for all allowed values of the parameter $b$.  We have considered the evolution of the scalar curvature in a spherically collapsing object. The numerical solution of the Eq.~\eqref{dyneqiny} is plotted in Fig.~\ref{fig:dynsolution}. From the plot, it is evident that the time taken to reach the singularity is finite and much less than the age of the universe. 

In conclusion, the model of Eq~.\eqref{model} is plagued with the issue of fatal curvature singularity. Additionally, we also have found that the model does not satisfy the fifth force constraint, and therefore it is not a viable model for the late time Universe showing acceleration. 

\subsection*{Acknowledgements}
K Dutta would like to acknowledge support from a Ramanujan fellowship of the Department of Science and Technology, India and Max Planck Society-DST Visiting Fellowship grant.

\end{document}